\documentclass[photonics,communication,accept,pdftex,moreauthors]{Definitions/mdpi} 
\firstpage{1} 
\pubvolume{1}
\issuenum{1}
\articlenumber{0}
\pubyear{2025}
\copyrightyear{2025}
\datereceived{30 March 2025} 
\daterevised{21 April 2025} 
\dateaccepted{23 April 2025}
\datepublished{ } 
\hreflink{https://doi.org/} 

\usepackage{braket}
\Title{Asymmetry Analysis of the Autler--Townes Doublet in the Trap-Loss Fluorescence Spectroscopy of Cesium MOT with Single-Step Rydberg Excitation}

\TitleCitation{Asymmetry Analysis of the Autler--Townes Doublet in the Trap-Loss Fluorescence Spectroscopy of Cesium MOT with Single-Step Rydberg Excitation}

\Author{Xiaokai Hou $^{1}$, Yuewei Wang $^{1}$, Jun He $^{1,2}$ and Junmin Wang $^{1,2,}$*\orcidA{}}

\AuthorNames{Xiaokai Hou, Yuewei Wang, Jun He and Junmin Wang}

        \AuthorCitation{Hou, X.; Wang, Y.; He, J.; Wang, J.}

\address{%
$^{1}$ \quad State Key Laboratory of Quantum Optics Technologies and Devices, Institute of Opto-Electronics,\linebreak Shanxi University,
Taiyuan 030006, China; 202112607006@email.sxu.edu.cn (X.H.); 202312607025@email.sxu.edu.cn (Y.W.); hejun@sxu.edu.cn (J.H.)
\\
$^{2}$ \quad Collaborative Innovation Center of Extreme Optics, Shanxi University, Taiyuan 030006, China}

\corres{Correspondence: wwjjmm@sxu.edu.cn}

\abstract{The Autler--Townes (AT) doublet, a fundamental manifestation of quantum interference effects, serves as a critical tool for studying the dynamic behavior of Rydberg atoms.
Here, we investigate the asymmetry of the Autler--Townes (AT) doublet in the trap-loss fluorescence spectroscopy (TLFS) of cesium (Cs) atoms confined in a magneto-optical trap (MOT) with single-step Rydberg excitation using a 319-nm ultraviolet (UV) laser. A V-type three-level system involving the ground state $6\text{S}_{1/2}$ ($\text{F}$ = 4), excited state $6\text{P}_{3/2}$ ($\text{F}^{'}$ = 5), and Rydberg state ($n\text{P}_{3/2}$ ($\text{m}_\text{J}$ = +3/2)) is theoretically modeled to analyze the nonlinear dependence of the AT doublet's  
asymmetry and interval on the cooling laser's detuning. Experiments reveal that as the cooling laser detuning $\Delta_1$ decreases from $-$15~MHz to $-$10~MHz, the AT doublet exhibits increasing symmetry, while its interval shows a nonlinear decrease. Theoretical simulations based on the density matrix equation and Lindblad master equation align closely with experimental data, confirming the model's validity. This study provides insights into quantum interference dynamics in multi-level systems and offers a systematic approach for optimizing precision measurements in cold atom~spectroscopy.}

\keyword{single-step Rydberg excitation; trap-loss fluorescence spectroscopy} 

\begin{document}


\section{Introduction}
 Rydberg atoms \cite{gallagher2005rydberg}, highly-excited atoms with principal quantum numbers n $\textgreater$ 10, have shown great promise in terms of their unique structure. These physical properties scale with principal quantum number n, including long radiation lifetimes (${\sim}n^3$), strong dipole–dipole interactions (${\sim}n^4$), and large polarizability (${\sim}n^7$). They have found extensive applications in many-body physics \cite{browaeys2020many}, quantum information and computing \cite{adams2019rydberg,saffman2010quantum, Phuttitarn2024Enhanced}, and precision measurement \cite{arias2019realization,jing2020atomic,bai2019single,anderson2016optical,kumar2017rydberg, Meng2023Machine}. To exploit coherent quantum 
dynamics, these experiments are performed on timescales 
shorter than the lifetime of the Rydberg state. However, for the 
realization of supersolids \cite{Boninsegni2012Colloquium}, frustrated quantum magnetism \cite{Glaetzle2015Designing} or spin squeezing for enhanced metrology \cite{Gil2014Spin}, it is 
necessary to extend the investigation time of Rydberg atoms. To take advantage of the long coherence time of ground-state atoms and the strong long-range interaction between  Rydberg atoms, we can create a wave function that is mostly 
ground state with an adjustable Rydberg component by 
off-resonantly coupling the ground state to the Rydberg state 
\cite{arias2019realization,Lee2017Demonstration,Jau2016Entangling}, which is called the Rydberg dressing approach. Therefore, experimental research on the Autler–Townes (AT) doublet \cite{autler1955stark,Zhang2013Autler, Ahmed2011Quantum} has opened a route to produce the dressed-state atoms with the long coherence time and controllable long-range interaction by adjusting the coupling intensity and detuning between the ground state and Rydberg~state.  

\textls[-35]{The AT doublet, a quintessential manifestation of quantum interference effects, is characterized by key properties such as the interval, symmetry and the width of the spectral double-peaks structure, both of which are intricately linked to parameters like the Rabi frequency and detuning of the coupling laser field \cite{bai2019single,wang2023autler,cao2022dephasing}. Consequently, a comprehensive investigation of AT doublet properties is pivotal for elucidating the dynamic mechanisms of the Rydberg-dressing ground state and for uncovering the nonlinear effects in complex systems.}

In this paper, we systematically investigate the characteristics of the AT doublet of the trap-loss fluorescence spectroscopy (TLFS) of in cesium (Cs) magneto-optical traps (MOT) using single-step ultraviolet (UV) laser excitation. First, a theoretical model describing the three-level V-type system of Cs atoms is established, analyzing the relationship between cooling laser detuning and AT doublet characteristics. Second, using TLFS, experimental observations of the AT doublet under different cooling laser parameters are conducted, confirming that the symmetry of the AT doublet improves as detuning decreases, while the AT doublet interval exhibits a nonlinear reduction trend.

\section{The V-Type Three-Level System with Rydberg State of Cs Atoms}
\begin{figure}[H]
    
    \hspace{-3pt}\includegraphics[width=0.6\textwidth,height=0.33\textwidth]{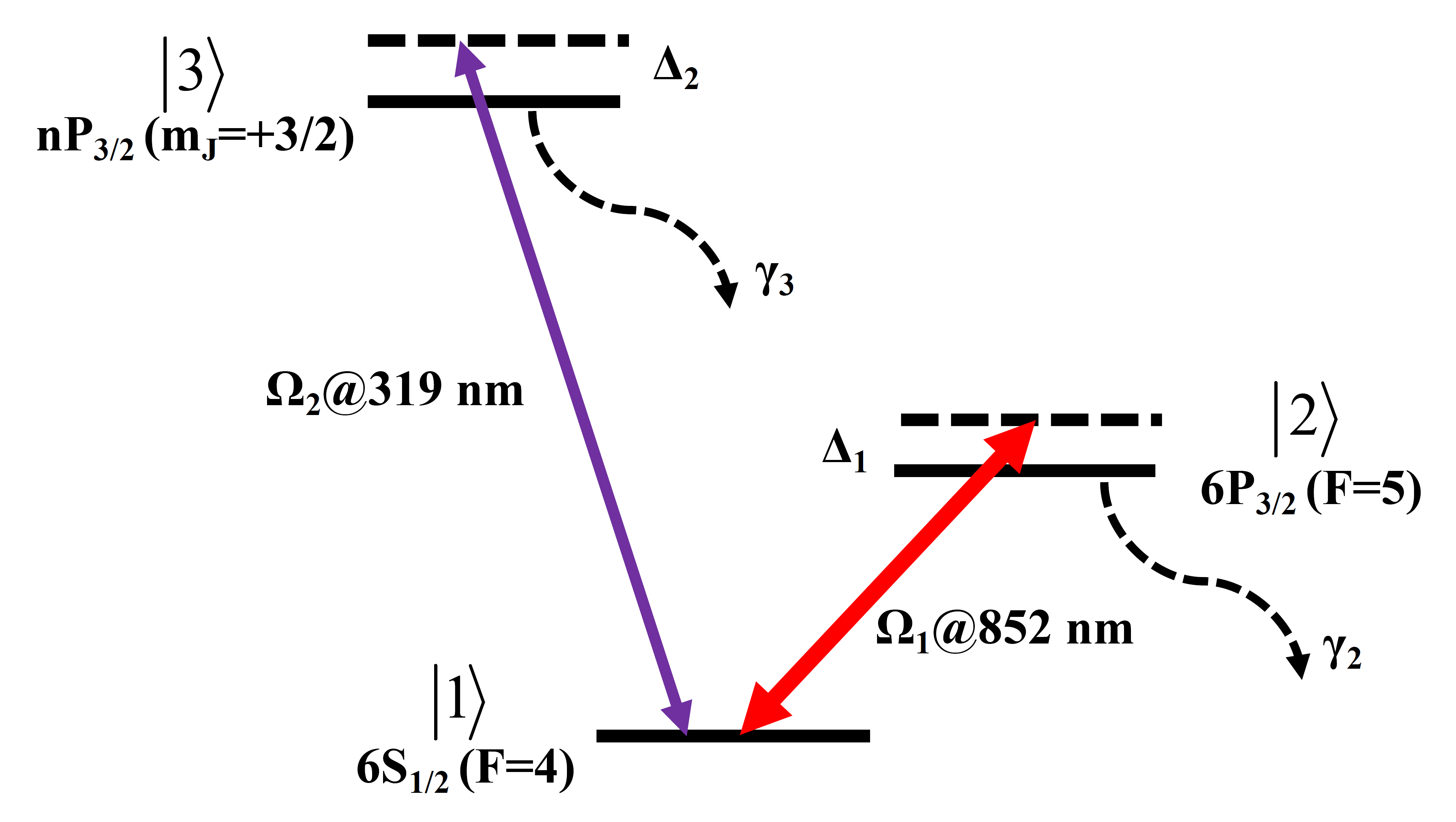}
    \caption {Schematic
 diagram of the V-type three-level system of Cs atoms. State $\ket{1}$ ($6\text{S}_{1/2}$ (F~=~4)) and state $\ket{2}$ ($6\text{P}_{3/2}$ ($\text{F}^{'}$ = 5)) are coupled by an infrared laser ($\Omega_1$@852 nm) with detuning $\Delta_1$ and spontaneous decay rate $\gamma_{2}$. Simultaneously, the UV laser ($\Omega_2$@319 nm) excites Cs atoms from state $\ket{1}$ to state $\ket{3}$ ($n\text{P}_{3/2}$ ($\text{m}_\text{J}$ = +3/2)) with corresponding detuning $\Delta_2$ and spontaneous decay rate $\gamma_{3}$. Dashed arrows indicate radiation decay channels, while solid arrows represent laser coupling~pathways.}
    \label{fig:1}
\end{figure}

\textls[-15]{As depicted in Figure \ref{fig:1}, here is 
a V-type three-level system \cite{Zhang2004asymmetry} comprising the Cs ground state $6\text{S}_{1/2}$ (F = 4) (denoted as $\ket{1}$), the Cs atom excited state $6\text{P}_{3/2}$ ($\text{F}^{'} = 5$) (denoted as $\ket{2}$), and the Cs atom Rydberg state $n\text{P}_{3/2}$ ($\text{m}_\text{J}$ = +3/2) (denoted as $\ket{3}$). $\Omega_1$ denotes the Rabi frequency of the excitation laser (coupling Levels $\ket{1}$ and $\ket{2}$); $\Delta_1$ represents the detuning of the excitation laser; $\gamma_{2}$ represents spontaneous decay rate of the excitation state; $\Omega_2$ denotes the Rabi frequency of the UV laser (coupling Levels $\ket{1}$ and $\ket{3}$); and $\Delta_2$ represents the UV laser detuning, $\gamma_{3}$ represents spontaneous decay rate of the Rydberg state $\ket{3}$.}

In order to explain the dynamics of the interaction between light and atoms in a V-type three-level system, we consider the following theory,

\begin{equation}
H=\frac \hbar2\begin{pmatrix}0&\Omega_1&\Omega_2\\\Omega_1&-2\Delta_1&0\\\Omega_2&0&-2\Delta_2\end{pmatrix}
\end{equation}

\begin{equation}
\rho=\begin{pmatrix}\rho_{11}&\rho_{12}&\rho_{13}\\\rho_{21}&\rho_{22}&\rho_{23}\\\rho_{31}&\rho_{32}&\rho_{33}\end{pmatrix}
\end{equation}

\begin{equation}
    \left.L(\rho)=\left(\begin{array}{ccc}\gamma_2\rho_{22}+\gamma_3\rho_{33}&-\frac12\gamma_2\rho_{12}&-\frac12\gamma_3\rho_{13}\\\\-\frac12\gamma_2\rho_{21}&-\gamma_2\rho_{22}&-\frac12\left(\gamma_2+\gamma_3\right)\rho_{23}\\\\-\frac12\gamma_3\rho_{31}&-\frac12\left(\gamma_2+\gamma_3\right)\rho_{32}&-\gamma_3\rho_{33}\end{array}\right.\right)
\end{equation}

\begin{equation}
    \dot{\rho}=-\frac i\hbar[H,\rho]+L(\rho)
\end{equation}

 Here, $H$ represents the Hamiltonian describing the interaction of the V-type three-energy-level atoms with the laser field; $\rho_{ij} (i,j=1,2,3)$ denotes the elements of the density matrix. When $i=j$, $\rho_{ii}$ represents the diagonal elements of the density matrix, corresponding to the atomic population probabilities of each energy level. Conversely, when $i\neq j$, $\rho_{ij}$ describes the non-diagonal elements, characterizing the coherence between two energy levels.  $L(\rho)$ represents the decoherence matrix, which accounts for the dissipation processes within the system. $\gamma_2$ represents the spontaneous radiation decay rate of the cesium atom $6\text{P}_{3/2}$; meanwhile, $\gamma_3$ denotes the spontaneous radiation decay rate of the cesium atom Rydberg state $n\text{P}_{3/2}$.

The TLFS \cite{wang2023autler,cao2022dephasing,halter2022trap} serves as a powerful tool to characterize the dynamic process of cold-atom fluorescence attenuation induced by Rydberg excitation within a MOT. In this context, cold-atom fluorescence arises from the spontaneous radiation of the excited state, rendering it directly proportional to the atomic population probability of the excited state. Utilizing Equations (1)--(4), we derive the instantaneous steady-state solution, incorporating the temporal density matrix to analyze the population probability $\rho_{22}$ of energy level 2.

\begin{figure}[H]
 
    \includegraphics[width=0.93\textwidth,height=0.41\textwidth]{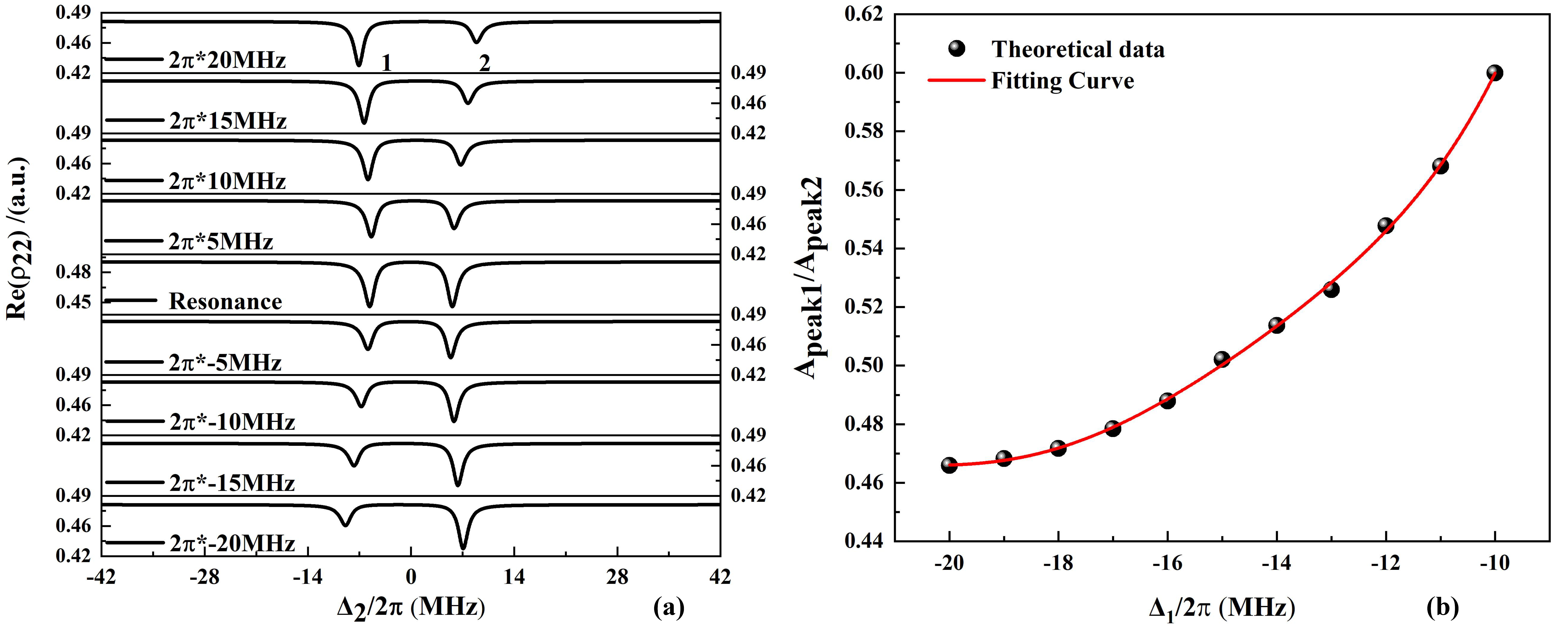}
    \caption{(\textbf{a}) According to Equations (1)--(4), simulations are carried out to obtain the relationship between Re($\rho_{22}$) and UV laser detuning $\Delta_2$, which exhibits an obvious AT bimodal structure; the detuning of the excitation laser $\Delta_1$ is constantly changed. When detuning is gradually reduced, the AT doublet tends to be gradually symmetric. (\textbf{b}) For the purpose of our actual experiments (the cooling laser frequency to maintain the normal operation of the MOT should be negatively detuned), it is taken as $-$20 MHz to $-$10 MHz, and the ratio of their respective corresponding double-peak amplitudes. The black squares are simulated data points.The fitting curve is a 5th-order polynomial, the expression is $\mathbf{y=Ax^5+Bx^4+Cx^3+Dx^2+Ex+F}$, and the COD is 0.9993. Among them, the parameters are as follows: A = $3.13 \times 10^{-6}$, B = $2.56 \times 10^{-4}$, C = $8.320 \times 10^{-3}$, D = 0.1341, E = 1.09, and F = 4.149.}
    \label{fig:2}
\end{figure}

As depicted in Figure \ref{fig:2}, the excitation laser Rabi frequency $\Omega_1$ and the UV laser Rabi frequency $\Omega_2$ are maintained constant, while UV laser detuning $\Delta_2$ is scanned  
 to generate the double-peak spectral structure. Since the vertical axis represents the atomic population probability of excited state 2, which is directly proportional to the fluorescence intensity of the cold atoms, this simulation effectively characterizes the physical behavior of cold atoms in a MOT with reduced fluorescence due to Rydberg excitation. In this spectrum, Peaks 1 and 2 correspond to the double peaks of the AT doublet. These peaks exhibit increased symmetry as the excitation laser detuning approaches resonance, while the interval between the peaks expands with increasing excitation laser detuning. We fit the data using a fifth-order polynomial with an expression $\mathbf{y=Ax^5+Bx^4+Cx^3+Dx^2+Ex+F}$
 and the coefficient of determination (COD) of 0.9993. Among them, the parameters are as follows: A = $3.13 \times 10^{-6}$, B = $2.56 \times 10^{-4}$, C = $8.320 \times 10^{-3}$, D = 0.1341, E = 1.09, and F = 4.149.

\section{Trap-Loss Fluorescence Spectroscopy and AT Doublet in Cs MOT}
\begin{figure}[H]
    
    \includegraphics[width=0.95\textwidth,height=0.53\textwidth]{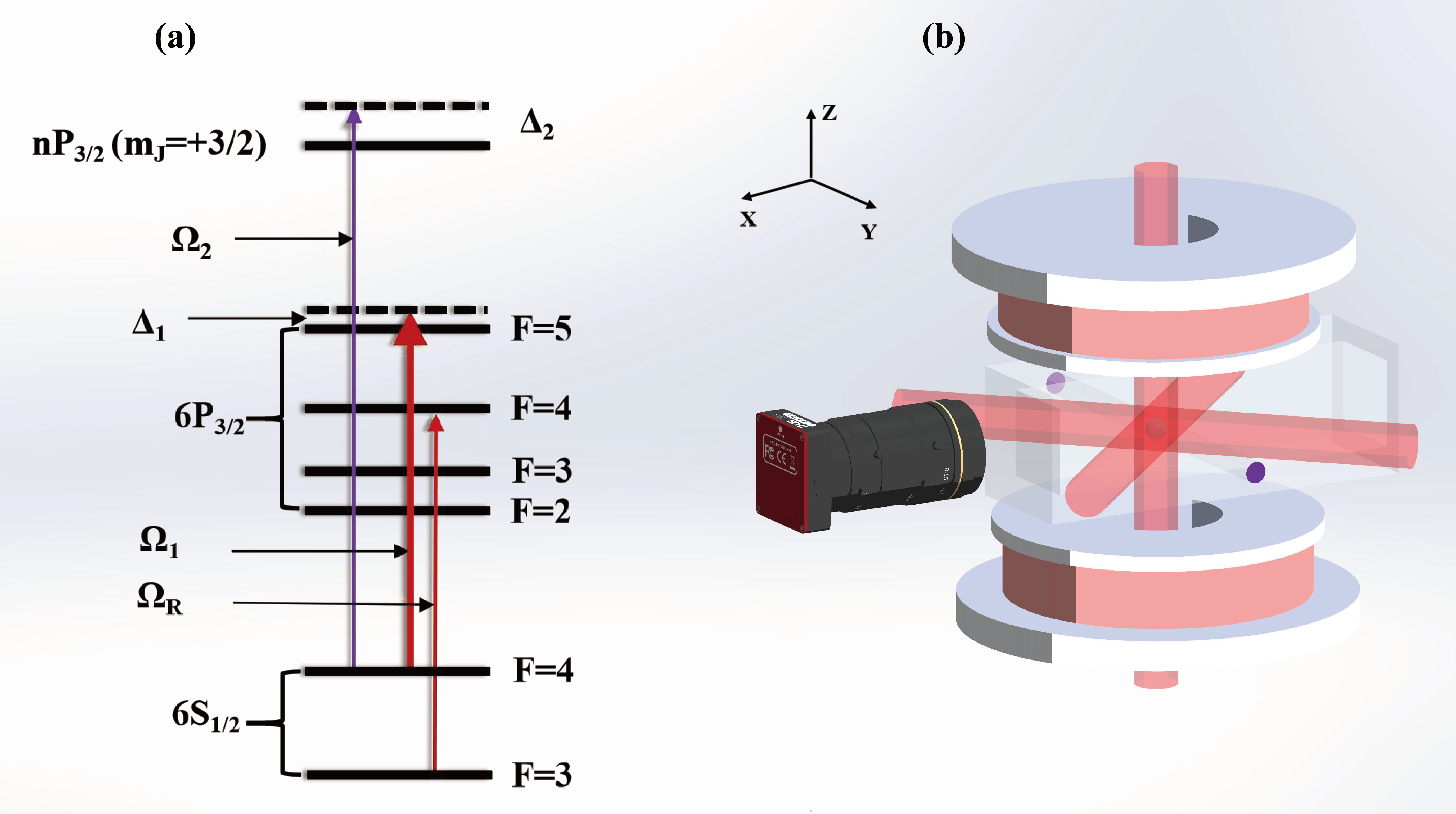}
    \caption{Diagram of the Cs relevant energy level and MOT setup. (\textbf{a}) Atomic energy level diagram, emphasizing the cooling laser beam (thick red arrows), repumping laser beam (thin red arrows), and UV laser beam (purple arrows) for single-step Rydberg excitation. (\textbf{b}) Experimental configuration, featuring the trapping region surrounded by a pair of coils generating a quadrupole magnetic field and the intersection of laser beams for cooling. The imaging system, aligned along a principal axis, enables the precise monitoring of atomic fluorescence.}
    \label{fig:3}
\end{figure}

As illustrated in Figure \ref{fig:3}, 
the image on the left shows the atomic energy level diagram of the MOT and the single-step Rydberg excitation. The thick red arrow represents the cooling laser beam, its Rabi frequency is marked as $\Omega_{1}$, and the detuning is $\Delta_{1}$. The thin red arrow represents the repumping laser beam, which is used to pump the atoms populated on 6$\text{S}_{1/2}$ (F = 3) to continue to participate in the cooling process. The purple arrow represents the UV laser beam for single-step Rydberg excitation, with a Rabi frequency labeled as $\Omega_{2}$ and a detuned frequency as $\Delta_{2}$. The above marks correspond one-to-one to Figure \ref{fig:1} in the theoretical part; however, we ignore the possibility that the atoms will populate in 6$\text{S}_{1/2}$ (F = 3), i.e., the repumping laser is not considered. 
Furthermore, in Figure \ref{fig:3}b, the Cs MOT is housed within a 30 $\times$ 30 $\times$ 120 $\text{mm}^{3}$ ultra-high vacuum (UHV) glass cell, with typical pressure of approximately 1 $\times$ $10^{-9}$ Torr. The frequency of the cooling laser beam, which has a Gaussian diameter of approximately 10 mm, is locked to the Cs $6\text{S}_{1/2}$ (F = 4) $\rightarrow$  $6\text{P}_{3/2}$ ($\text{F}$ = 5) cycling transition with  detuning $\Delta_{1}$. The repumping laser beam, with a Gaussian diameter of around 9 mm, is resonant with the Cs $6\text{S}_{1/2}$ (F = 3) $\rightarrow$ $6\text{P}_{3/2}$ ($\text{F}$ = 4) transition. The cooling laser beam and repumping laser beams are combined and are then divided into three parts as follows: one part along the Z and $-$Z directions, accounting for approximately 40$\%$ of the power; the remaining two parts are split equally in power and directed in the XY plane at an angle of about $80^{\circ}$. A gradient magnetic field is produced by a pair of anti-Helmholtz coils driven by a constant-current source, generating a typical axial magnetic field gradient of 10 Gauss/cm with a current of 9 A. Consequently, a bright cloud of cold atomic ensemble can be immediately observed using a CCD camera. 

Since a MOT cannot trap Rydberg atoms, Rydberg atoms are lost when the UV\linebreak laser~\cite{wang2016Realization,wang2016development,bai2017Electronic} couples Cs atoms from the $6\text{S}_{1/2}$ (F = 4) to the $n$$\text{P}_{3/2}$ ($\text{m}_{J}$ = +3/2) state, achieving single-step Rydberg excitation. In other words, the excitation of Rydberg atoms can be inferred by observing the trap-loss fluorescence from a cold cloud in the MOT. 

 During the experiment, the change in frequency of our UV laser system is achieved by scanning it point by point. Each frequency point lasts for 30 s. The CCD camera is triggered for taking the photograph of the cold atom within 30 s, the timestamp of the initial trigger is synchronized, which occurs 5 s after the UV laser is adjusted to the target frequency, and taking the photograph ends at the next time the UV laser frequency is modified, the exposure time of the CCD camera is 5 ms and the CCD’s trigger frequency is 0.2 Hz. In a spectrum acquisition process, each frequency point lasts for 30 s, and the UV laser frequency needs to be changed about 45$\sim$55 times, covering about 175$\sim$200 MHz, with a total duration of about 30 min.

Since the UV Rabi frequency is in the order of several hundred kHz, the Rydberg excitation time must be less than the CCD exposure time. Therefore, we believe that the loss of atomic number in the MOT is at a completely steady state within the 5 ms time window. Within 30 s, five fluorescence images of cold atoms will be obtained, and three of them will be randomly selected to calibrate the remaining proportion of cold atoms in the MOT at the current UV laser frequency. For every five data acquisitions, we will take pictures of the MOT that are not Rydberg excited by UV laser for image processing and fluorescence intensity normalization for these five data acquisitions. After the spectral data acquisition, we use a MATLAB R2020a program to process the fluorescence images of cold atoms, and the relative fluorescence intensity is calculated by dividing the total fluorescence intensity of the cold atoms in the MOT with UV laser excitation by the reference fluorescence intensity without UV laser excitation. In addition, background light was subtracted to correct for the effect of background scattered light.

Throughout the experiment, we utilize a wavelength meter (Germany, HighFinesse, WS-7)
 to monitor the red light frequency in real time, ensuring the accuracy of the UV frequency. Additionally, the wave meter is calibrated after every five acquisitions using the probe beam ($6\text{S}_{1/2}$ (F = 4) $\rightarrow$ $6\text{P}_{3/2}$ ($\text{F}$ = 5), with 351.72196 THz \cite{Kramida2024nist}) as the frequency reference.
 
\begin{figure}[H]
 
    \includegraphics[width=0.95\textwidth,height=0.72\textwidth]{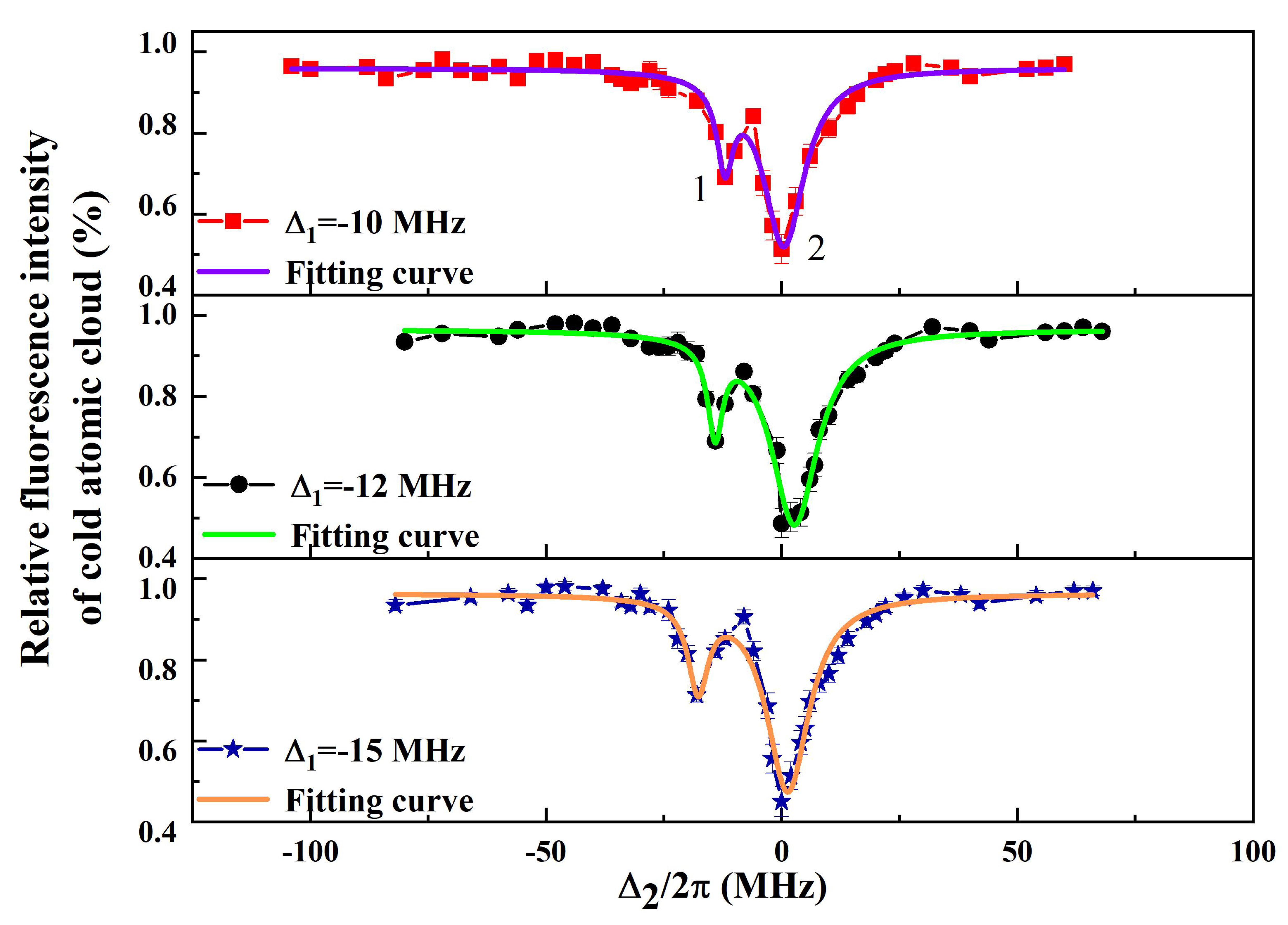}
    \caption{TLFS and their AT doublet at different levels of cooling laser detuning. By adjusting the parameters, a significant change in the resonance signal can be observed. Among them, the cooling laser Rabi frequency is fixed to $2\pi\times 8.1$ MHz, the Rabi frequency of the UV laser is \mbox{$2\pi\times 117$ kHz}, and the detuning of the cooling laser is gradually reduced, with three sets of spectral signals ((\textbf{top}): $-$10~MHz; (\textbf{Medium}): $-$12 MHz; (\textbf{bottom}): $-$15 MHz). The amplitude ratio of Peaks 1 and 2 gradually increases and tends to be asymmetrical, and the AT doublet interval also increases gradually with the increase in cooling laser detuning. The red squares, black circles, and blue stars represent the experimental data, and the purple, green, and orange curves are the multi-peak Lorentzian profile fitting curves.}
    \label{fig:4}
\end{figure}

\textls[-15]{We select $71\text{P}_{3/2}$ as the target Rydberg state, experimentally observe the TLFS and its AT doublet, and investigate the dependence of AT doublet characteristics on laser detuning. Specific results are given in \cite{bai2019single,wang2023autler} In this paper, we only analyze the symmetry and interval of the AT doublet of the TLFS. As shown in Figure \ref{fig:4}, we fixed the Rabi frequency $\Omega_{1}$ of the cooling laser beam to $2\pi\times 8.1$ MHz and the UV laser Rabi frequency $\Omega_{2}$ to \mbox{$2\pi\times 117$ kHz}, and we gradually decreased the cooling laser detuning $\Delta_{1}$ ($-$15, $-$14, $-$13, $-$12, $-$11, $-$10~MHz). Here, we present only three of these sets of data ((top): $-$10 MHz; (middle): $-$12 MHz; (bottom): $-$15 MHz). The ratio of the amplitudes of the AT doublet (Peaks 1 and 2) gradually decreases with the increase in cooling laser detuning, i.e., the double peaks gradually tend to be asymmetric, and according to $\tilde{\Omega}=\sqrt{\Omega_{1}^2+\Delta_{1}^2}$, the intervals of the AT doublet gradually decrease with the decrease in detuning.}
\vspace{-6pt}

However, to our surprise, the full width at half maximum (FWHM) of the TLFS measured at four different principal quantum numbers was in the order of 20 MHz in its entirety. We propose the following five possible mechanisms for the linewidth broadening of the TLFS: 

Firstly, the Rabi frequency of the cooling laser is still at a relatively high order of magnitude, which still causes the ground state to be broadened by a factor of 1.7 due to power broadening. Secondly,  Zeeman broadening in the z-direction at the center of the MOT can be calculated according to $\Delta\Gamma=\frac{\mu_B}{h}g_J(S_{1/2})\frac{\partial B}{\partial z}\Delta z$. The ground state $6\text{S}_{1/2}$ broadens to 0.96 MHz due to the presence of a magnetic field. Thirdly, our atoms are exposed to severe blackbody radiation (BBR) at room temperature
that will directly couple the target Rydberg state with the adjacent Rydberg state. So, the lifetime of the target Rydberg state decreases dramatically, the typical Rydberg state lifetime is about 10 $\upmu$s, and the width of the Rydberg state energy level is broadened to 100 kHz. Fourth, as can be seen in Figure~\ref{fig:4}, the baseline is less than 100$\%$ at the UV laser separation resonance position, and there is still a $15\%{\sim}20\%$ loss of cold atomic fluorescence intensity. This is due to the fact that the cold atoms in the magneto-optical trap are constantly affected by ionization, and we speculate that the mechanisms that lead to ionization are photoionization, field ionization, and collision ionization. Due to the presence of ionization, the overall spectral signal increases by $15\%{\sim}20\%$, and then, we can simply infer that the spectral linewidth is broadened by at least 15$\%$ due to the presence of ionization. Last, due to the existence of the ionization mechanism, a cold plasma is formed near the cold atoms, which in turn generates a local electric field of a certain intensity, whose electric field intensity varies spatially. When such a local electric field is superimposed with the background DC electric field, the Stark shift formed on the Rydberg state will be spatially dependent. It can be roughly inferred that the remaining broadening of about 5 MHz comes from the local electric field generated by the cold plasma \cite{Robinson2000Spontaneous}.

\begin{figure}[H]
  
\includegraphics[width=0.95\textwidth,height=0.41\textwidth]{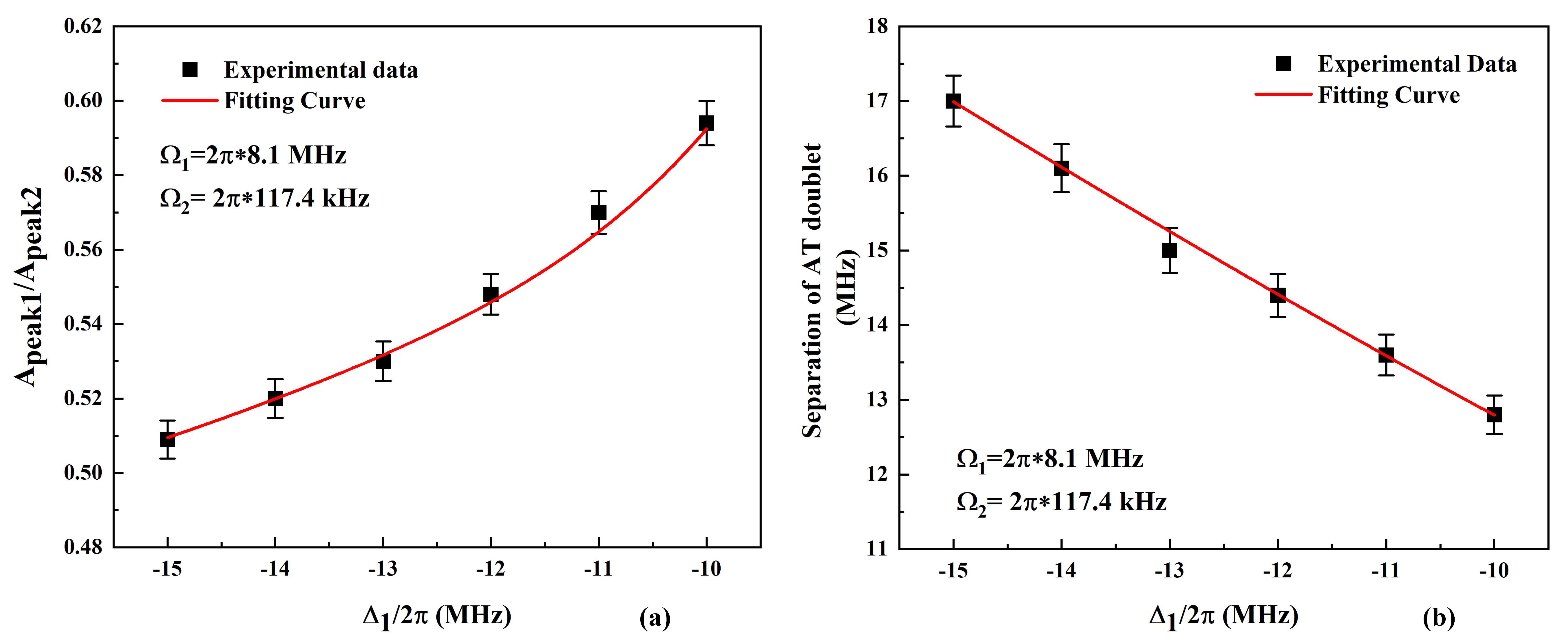}
    \caption{Comparative analysis of experimental results obtained under varying laser detuning \linebreak  ($-$15, $-$14, $-$13, $-$12, $-$11, and $-$10 MHz). (\textbf{a}) The black data points illustrate the trend of the AT doublet amplitude ratio as a function of cooling laser detuning, with uncertainties indicated by error bars. (\textbf{b}) The black data points represent the AT doublet interval, while the nonlinear dependence predicted by the theoretical model is highlighted through the red solid fitting line.}
    \label{fig:5}
\end{figure}

Figure \ref{fig:5}a illustrates the relationship between AT doublet symmetry and cooling laser detuning experimentally, and we select the range of laser detuning of $-$15$\sim$$-$10 MHz with a step of 1 MHz. It can be seen that the variation trend of the black data points is consistent with that of Figure \ref{fig:2}b, and the AT doublet gradually tends to be symmetrical as the amount of cooling laser detuning decreases. The red curve is the fitting function, which is still a fifth-order polynomial, and the parameters remain unchanged, with a COD of 0.9451. This shows that the theoretical model is in good agreement with the experimental data. Figure~\ref{fig:5}b shows the AT doublet splitting interval with cooling laser detuning in the three sets of spectra. With the decrease in cooling laser detuning, the double-peak interval tends to decrease, and we use the formula with the expression $\tilde{\Omega}=A\sqrt{\Omega_{1}^2+\Delta_{1}^2}$, where A is the correction parameter, the specific value is 0.91, and the COD is 0.9997.

\section{Conclusions}
 In this paper, we systematically analyze the behavior of AT doublet asymmetry in the MOT of Cs atom through theoretical modeling and experiments. We used a 319~nm UV laser to excite the ground-state atoms $6\text{S}_{1/2}$ (F = 4) to the Rydberg state $71\text{P}_{3/2}$ ($\text{m}_\text{J}$~=~3/2) by a single-step Rydberg excitation, and we obtained the TLFS by all-optical non-destructive detection. Due to the presence of a strong cooling laser, we also clearly observed the AT doublet of the TLFS and comprehensively analyzed the relationship between the asymmetry of the AT doublet and the detuning of the cooling laser by establishing a V-type three-level system model. 
 
 In subsequent experiments, we will use a 319 nm UV laser system to achieve single-step Rydberg excitation of the cesium-cold atomic ensemble to achieve a Rydberg weakly dressing ground state by precisely controlling laser detuning, Rabi frequency, and polarization. This novel quantum resource combines the long coherence time of the ground state with the strong interaction of the Rydberg state, which can significantly improve the fidelity of quantum logic gates. So, the experimental study of the Autler--Townes (AT) splitting of single-step Rydberg excitation spectroscopy is pivotal for elucidating the dynamic mechanisms of the Rydberg dressing ground state and for uncovering nonlinear effects in complex systems.
  \vspace{6pt}

%

\begin{thebibliography}{999}
\bibitem{gallagher2005rydberg} Gallagher, T.F. \emph{Rydberg Atoms}; {Cambridge University Press:} Cambridge, UK, 2005.


\bibitem{browaeys2020many}
Browaeys, A.; Lahaye, T. Many-body physics with individually controlled {Rydberg} atoms. {\em Nat. Phys.} \textbf{2020}, \emph{16}, 132--142.

\bibitem{adams2019rydberg}
Adams, C.S.; Pritchard, J.D.; Shaffer, J.P. Rydberg atom quantum technologies. {\em J. Phys. B At. Mol. Opt. Phys.} \textbf{2019}, \emph{53}, 012002.

\bibitem{saffman2010quantum}
Saffman, M.; Walker, T.G.; M{\o}lmer, K. Quantum information with {Rydberg} atoms. {\em Rev. Mod. Phys.} \textbf{2010}, \emph{82}, 2313.

\bibitem{Phuttitarn2024Enhanced}
Phuttitarn, L.; Becker, M.; Chinnarasu, R.; Graham, M.; Saffman, M.  Enhanced measurement of neutral-atom qubits with machine learning. {\em Phys. Rev. Appl.} \textbf{2024}, \emph{22}, 024011.

\bibitem{arias2019realization}
Arias, A.; Lochead, G.; Wintermantel, T.M.; Helmrich, S.; Whitlock, S. Realization of a {Rydberg}-dressed {Ramsey} interferometer and electrometer. {\em Phys. Rev. Lett.} \textbf{2019}, \emph{122}, 053601.

\bibitem{jing2020atomic}
Jing, M.Y.; Hu, Y.; Ma, J.; Zhang, H.; Zhang, L.J.; Xiao, L.T.; Jia, S.T. Atomic superheterodyne receiver based on microwave-dressed {Rydberg} spectroscopy {\em Nat. Phys.} \textbf{2020}, \emph{16}, 911--915.

\bibitem{bai2019single}
Bai, J.D.; Liu, S.; Wang, J.Y.; He, J.; Wang, J.M. Single-photon {Rydberg} excitation and trap-loss spectroscopy of cold cesium atoms in a magneto-optical trap by using of a 319-nm ultraviolet laser system. {\em IEEE J. Sel. Top. Quantum Electron.} \textbf{2020}, \emph{26}, 1600106.

\bibitem{anderson2016optical}
Anderson, D.A.; Miller, S.A.; Raithel, G.; Gordon, J.A.; Butler, M.L.; Holloway, C.L. Optical measurements of strong microwave fields with {Rydberg} atoms in a vapor cell. {\em Phys. Rev. Appl.} \textbf{2016}, \emph{5}, 034003.

\bibitem{kumar2017rydberg}
Kumar, S.; Fan, H.Q.; K{\"u}bler, H.; Jahangiri, A.J.; Shaffer, J.P. {Rydberg}-atom based radio-frequency electrometry using frequency modulation spectroscopy in room temperature vapor cells. {\em Opt. Express} \textbf{2017}, \emph{25},~8625--8637.

\bibitem{Meng2023Machine}
Meng, X.; Zhang, Y.W.; Zhang, X.C.; Jin, S.C.; Wang, T.R.; Jiang, L.; Xiao, L.T.; Jia, S.T.; Xiao, Y.H. Machine learning assisted vector atomic magnetometry. {\em Nat. Commun.} \textbf{2023}, \emph{14}, 6105.

\bibitem{Boninsegni2012Colloquium}
Boninsegni, M.; Prokof’ev, N.V.
 Colloquium: Supersolids: What 
and where are they? {\em Rev. Mod. Phys.} \textbf{2012}, \emph{84}, 759.

\bibitem{Glaetzle2015Designing}
Glaetzle, A.W.; Dalmonte, M.; Nath, R.; Gross, C.; Bloch, I.; Zoller, P. Designing frustrated quantum magnets with laser dressed Rydberg atoms. {\em Phys. Rev. Lett.} \textbf{2015}, \emph{114}, 173002.

\bibitem{Gil2014Spin}
Gil, L.I.R.; Mukherjee, R.; Bridge, E.M.; Jones, M.P.A.; Pohl, T. Spin squeezing in a Rydberg lattice clock. {\em PHysical Rev. Lett.} \linebreak  \textbf{2014}, \emph{112}, 103601.

\bibitem{Lee2017Demonstration}
Lee, J.; Martin, M.J.; Jau, Y.Y.; Keating, T.; Deutsch, T.H.; Biedermann, G.W. Demonstration of the Jaynes--Cummings ladder with Rydberg-dressed atoms. {\em Phys. Rev. A} \textbf{2017}, \emph{95}, 041801(R).

\bibitem{Jau2016Entangling}
Jau, Y.Y.; Hankin, A.M.; Keating, T.; Deutsch, T.H.; Biedermann, G.W.  Entangling atomic spins with a Rydberg-dressed spin-flip blockade. {\em Nat. Phys.} \textbf{2016},  \emph{12}, 71--74.


\bibitem{autler1955stark}
Aulter, S.H.; Townes, C.H. Stark effect in rapidly varying fields. {\em Phys. Rev.} \textbf{1955}, \emph{100}, 703--723.

\bibitem{Zhang2013Autler}
Zhang, H.; Wang, L.M.; Chen, J.; Li, H.; Bao, S.X.; Zhang, L.J.; Zhao, J.M.; Jia, S.T.  Autler-Townes splitting of a cascade system in ultracold cesium {Rydberg} atoms. {\em Phys. Rev. A} \textbf{2013}, \emph{87}, 033835.

\bibitem{Ahmed2011Quantum}
Ahmed, E.H.; Ingram, S.; Kirova, T.; Salihoglu, O.; Huennekens, J.; Qi, J.; Guan, Y.; Lyyra, A.M. Quantum control of the spin--orbit interaction using the Autler-Townes effect. {\em Phys. Rev. Lett.} \textbf{2011}, \emph{107}, 163601.


\bibitem{wang2023autler}
Wang, X.; Hou, X.K.; Lu, F.F.; Chang, R.; Hao, L.L.; Su, W.J.; Bai, J.D.; He, J.; Wang, J.M. Autler--townes doublet in the trap-loss fluorescence spectroscopy due to single-step direct {Rydberg} excitation of cesium cold atomic ensemble. {\em AIP Adv.} \textbf{2023}, \emph{13}, 035126.

\bibitem{cao2022dephasing}
Cao, Y.F.; Yang, W.G.; Zhang, H.; Jing, M.Y.; Li, W.B.; Zhang, L.J.; Xiao, L.T.; Jia, S.T. Dephasing effect of {Rydberg} states on trap loss spectroscopy of cold atoms. {\em J. Opt. Soc. Am. B} \textbf{2022}, \emph{39}, 2032--2036.

\bibitem{Zhang2004asymmetry}
Zhang, L.S.; Feng, X.M.; Li, X.W.; Li, H.; Fu, G.S. The asymmetry of the Autler–Townes doublet in a three-level system. \linebreak  {\em Chin. Phys.} \textbf{2004}, \emph{13}, 348--352.

\bibitem{halter2022trap}
Halter, C.; Miethke, A.; Sillus, C.; Hegde, A.; G{\"o}erlitz, A. Trap-loss spectroscopy of {Rydberg} states in ytterbium. {\em J. Phys. B At. Mol. Opt. Phys.} \textbf{2023}, \emph{56}, 055001.

\bibitem{wang2016Realization}
Wang, J.Y.; Bai, J.D.; He, J.; Wang, J.M. Realization and characterization of single-frequency tunable 637.2 nm high-power laser. {\em  Opt. Commun.} \textbf{2016}, \emph{370}, 150--155.
 
\bibitem{wang2016development}
Wang, J.Y.; Bai, J.D.; He, J.; Wang, J.M. Development and characterization of a 2.2 {W} narrow-linewidth 318.6 nm ultraviolet laser. {\em J. Opt. Soc. Am. B} \textbf{2016}, \emph{33}, 2020--2025.

\bibitem{bai2017Electronic}
Bai, J.D.; Wang, J.Y.; He, J.; Wang, J.M. Electronic sideband locking of a broadly tunable 318.6 nm ultraviolet laser to an ultra-stable optical cavity. {\em J. Opt.} \textbf{2017}, \emph{19}, 045501.

\bibitem{Kramida2024nist}
Kramida, A.; Ralchenko, Y.; Reader, J.; NIST ASD Team. \emph{NIST Atomic Spectra Database}, Version 5.12; 2024.
 \linebreak  Available online: \url{https://physics.nist.gov/asd} (accessed on 12 November 2024).

\bibitem{Robinson2000Spontaneous}
Robinson, M.P.; Tolra, B.L.; Noel, M.W.; Gallagher, T.F.; Pillet, P.  Spontaneous evolution of {Rydberg} atoms into an ultracold plasma. {\em Phys. Rev. Lett.} \textbf{2000}, \emph{85}, 4466--4469.

\end{thebibliography}
\authorcontributions{Conceptualization, J.W. and X.H.; methodology, X.H.; software, X.H.; \linebreak  validation, X.H. and Y.W.; formal analysis, X.H.; investigation, X.H.; writing---original draft preparation, X.H.; writing---review and editing, X.H., J.W., and J.H.; visualization, X.X.; supervision,~J.W.; project administration, J.W.; funding acquisition, J.W. All authors have read and agreed to the published version of the manuscript.}

\funding{This research was funded by the National Key R $\&$ D Program of China (2021YFA1402002), the National Natural Science Foundation of China (12474483), and the Fundamental Research Program of Shanxi Province (202403021211013).}

\dataavailability{The original contributions presented in this study are included in the article. Further inquiries can be directed to the corresponding author.}

\conflictsofinterest{The authors declare no conflicts of interest.}




\begin{adjustwidth}{-\extralength}{0cm}

\reftitle{References}

\PublishersNote{}
\end{adjustwidth}
\end{document}